\documentclass[footinbib,a4paper,aps,amsmath,amssymb,showpacs,10pt,twocolumn,floatfix]{revtex4}
\usepackage{graphicx}
\usepackage{graphics,color}
\usepackage{umlaut}

\newcommand{\bs}{\;\;\;\;\;}
\newcommand{\sms}{\;\;}

\newcommand{\expc}{{\rm exp \mathcal C}}

\begin{document}

\title{New technique for replica symmetry breaking with application to the SK-model at and near $T=0$}
\author{M.J. Schmidt and R. Oppermann}
\affiliation{Institut f. Theoretische Physik, Universität Würzburg, Am Hubland, 97074 Würzburg, FRG}
\date{\today}

\begin{abstract}
We describe a novel method which allows the treatment of high orders of replica-symmetry-breaking (RSB) at low temperatures as well as at $T=0$ directly, without a need for approximations or scaling assumptions. It yields the low temperature order function $q(a,T)$ in the full range $0\leq a <\infty$ and is complete in the sense that all observables can be calculated from it. The behavior of some observables and the finite RSB theory itself is analyzed as one approaches continuous RSB. The validity and applicability of the traditional continuous formulation is then scrutinized and a new continuous RSB formulation is proposed.
\end{abstract}

\maketitle

\section{Introduction}

The ordered phase of spin-glass models \cite{binder-young} has gained much attention over the past three decades, but still there are many open questions even in models which were designed to possess simple 'mean-field-type' solutions. Paradigmatic for the difficulties encountered in the description of the ordered phase is the appearance of an ultrametric structure in the mean-field theory of spin glasses, known as Sherrington-Kirkpatrick (SK) model \cite{Sherrington-Kirkpatrick}. The correct treatment of the SK model involves a hierarchical scheme as introduced by Parisi \cite{ParisiRSB}, which only recently has been proven \cite{Talagrand} to be exact in the limit of an infinite number of hierarchical steps of replica symmetry breaking (RSB).

Though at temperatures right below the freezing transition $T_C$, a small number $\kappa$ of RSB steps is not a bad approximation, at temperatures $T\ll T_C$ the convergence of the results with respect to $\kappa$ becomes worse. Traditionally, the $\kappa=\infty$ limit is formulated by a continuous theory where the functional free energy $f[\tilde q(x)]$ is maximized with respect to the Parisi order function $\tilde q(x),\sms 0\leq x\leq 1$ \cite{binder-young,ParisiRSB}. This has later been expressed by a closed self-consistency form, involving a set of numerically solvable partial differential equations \cite{Sommers-Dupont,crisanti-rizzo}. The treatment of RSB at finite temperatures is well established, meanwhile. Near $T_C$, there are even analytical solutions available. At zero temperature, however, the proper form of the theory is still under discussion. The standard differential equations of the continuous theory and their initial conditions get singular in the zero temperature limit and the traditional formalism breaks down.

In the literature, one finds two main directions of approaching the zero temperature limit. Pankov proposed a beautifully simple scaling Ansatz directly at infinite RSB which becomes exact in a certain limit at zero temperature \cite{Pankov}. This limit, however, is far from being the only important region of the full zero temperature solution as we shall show later. On the other hand, our group has developed an exact finite RSB approach, which involves some advanced numerics when trying to reach orders of RSB, high enough to obtain a confident extrapolation $\kappa\rightarrow\infty$. However, our approach allows for a numerically exact description of the {\it full} SK-model with all its difficulties and features at $T=0$. The numerical results can be used to construct and test analytical approximations or simpler alternative theories which capture the essential physics while being generalizable to more complex physical situations \cite{prl07}.

The low temperature formulation, which is developed in the present work, directly identifies and analyzes the issues of the infinite RSB limit at zero temperature and resolves them by appropriate rescaling of auxiliary quantities. It is based on the idea of rescaled Parisi block size parameters \cite{ParisiRSB,oppermann-prb2000} $a_i=\beta m_i$ with $\beta=T^{-1}$. By this transformation, the non-trivial structure of the zero-temperature order function of the SK-model can be resolved. In the traditional formalism, the structure is concentrated and hidden at the point $x=0$, where $\tilde q(x)$ gets singular. Both, Pankov's and our approach,
investigate this point, but while the scaling approach is incomplete in the sense that it e.g. only accounts for the short time observables or that it cannot describe the situation at finite external fields correctly, our approach yields the full zero temperature solution.

The paper is organized as follows. In section II, we give a short review of the issues encountered when performing the zero temperature limit and show the connections between the two approaches mentioned above. In section III, we derive the non-trivial zero-temperature limit of the SK-model at finite RSB. In section IV, some results at low temperatures, obtained with the finite RSB formalism are shown. In section V, we discuss the $\kappa\rightarrow\infty$ limit of our theory.

\section{The $T=0$ limit of $\tilde q(x)$}

The Parisi order function $\tilde q(x),\sms x\in[0,1]$ which is the central quantity at finite temperatures becomes a constant function at zero temperature ($\lim_{T\rightarrow0}q(x)=1,\sms x>0$) with a singularity at $x=0$. Since typical observables as e.g. the internal energy $u$ involve integrals of the form $u=\beta\int_0^1 dx(1-\tilde q(x))^2$, one is tempted to absorb the divergent factor $\beta$ by a transformation $x\rightarrow a=\beta x$. The non-trivial zero temperature structure which was concentrated at $x=0$ in the Parisi order function and created the singularity there is then 'blown up' and resolved by the new order function $q(a)$. The transformation might remind the reader of PaT scaling \cite{PaT-scaling}. The idea here is different, though, as we don't seek for a universal form of the order function (which is known to be nonexistent) but for a non-singular zero temperature theory.

\begin{figure}
\centering
\includegraphics[width=230pt]{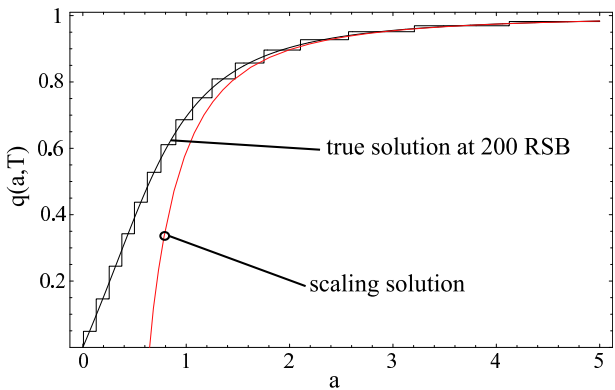}
\caption{(color-online): The true $T=0$ order function $q(a,0)$ at 200 RSB (black) and the step approximation at 20 RSB. The lower (red) curve shows the result of Pankov-scaling, i.e. $q(a,0) = 1-\gamma a^{-2}$.}
\label{pankov}
\end{figure}

The transformation $x\rightarrow a$ is not limited to $T=0$, of course, so that we can formulate a general low-temperature theory which allows a smooth transition from a treatment directly at $T=0$ to finite temperatures. The objective of our formalism is to find the function $q(a,T)$ with $a\in[0,\beta]$ from which one can obtain $\tilde q(x,T)=q(\beta x,T)$ at finite temperatures. At $\kappa<\infty$, $q(a,T)$ can be displayed as a step function with the number of steps equal to $\kappa$ (see Fig. \ref{pankov}).

In this notation, the Pankov-scaling approach is equivalent to expanding $q(a,0) \simeq 1-\gamma a^{-2}$ near $a=\infty$ and calculating the coefficient $\gamma$. The knowledge of the $a^{-2}$-term of this expansion is sufficient for short-time observables as the entropy or the non-equilibrium susceptibility. The full solution, however, must respect the functional dependence of $q(a,0)$ at all $a\in[0,\infty]$ which is needed to calculate e.g. the ground state energy or the equilibrium susceptibility. Another point is, that the action of a finite external field is reflected by the small $a$ domain of $q(a,T)$ which is not resolved in \cite{Pankov}. In Fig. \ref{pankov} the difference between the two solutions at low $a$ is shown.

In order to obtain a well defined zero temperature limit, we start with a formulation of the theory at finite orders of RSB ($\kappa<\infty$), where $q(a,T)$ is a function with $\kappa$ plateaus. The height of the $i$th plateau is given by the nunber $q_i \in [0,1]$ and the step positions are given by $a_i\in[0,\beta]$. If one aims at a continuous theory for $\kappa\rightarrow\infty$, one must first check that $\delta q_i = q_i - q_{i+1}$ and $\delta a_i = a_i - a_{i+1}$, i.e. the step heights and widths approach zero. For $T>0$, those requirements are fullfilled and are the basis of the well known Parisi differential equations \cite{ParisiRSB,binder-young,Sommers-Dupont}. At zero temperature and $\kappa\rightarrow\infty$, we still find $\delta q_i\rightarrow0$ which is not surprising, since $q_i$ is restricted to [0,1] independent of temperature. However, some step widths $\delta a_i$ grow indefinitely in this limit, so the assertion $\delta a_i\rightarrow 0$ which is needed to derive the Parisi differential equations fails at zero temperature. The non-vanishing $\delta a_i$ correspond to those $a_i$ which approach infinity for $\kappa\rightarrow\infty$ and so the continuous theory fails at $a=\infty$, while for finite $a$, a differential equation formulation is valid.

In section V of this paper, we show how the failure of a continuous theory at $a=\infty$ can be overcome by changing the initial condition of a partial differential equation.

\section{Recursion Technique for arbitrarily high RSB orders $\kappa$}
We investigate the standard SK-model Hamiltonian \cite{Sherrington-Kirkpatrick,binder-young} of $N$ Ising spins $s_i=\pm1$ with external field $H$
\begin{equation}
\mathcal H_{SK}=\sum_{i<j}J_{ij}s_is_j+H\sum_i s_i.
\end{equation}
The quenched random coupling constants $J_{ij}$ are Gaussian distributed random variables with zero mean and variance $N^{-1}$\footnote{The variance of the disorder distribution must scale as $N^{-1}$ in order to obtain a non-trivial thermodynamic limit. This choice also defines the energy and temperature scale of the system.}. The model yields a freezing transition at $T_C=1$. Since the coupling constants' degrees of freedom are quenched, the free energy per spin of the system is given by $f=-T/N\left<\log\left< \exp(-\beta\mathcal H_{SK})\right>_s\right>_J$ where $\left<\cdot\right>_s$ refers to an average in the space of Ising-spins $s_i=\pm 1$ and $\left<\cdot\right>_J$ refers to an average with respect to the coupling constants $J_{ij}$. In order to bypass the average of the logarithm we avail ourselves of the standard replica-trick. After transformation to a single site model \footnote{Due to the infinite range coupling between the spins, the effective dimension of the problem is $\infty$ and so the mean-field theory is exact}, we introduce Parisi RSB for the replica coupling matrix, perform the replica limit $n\rightarrow0$ and the thermodynamic limit $N\rightarrow\infty$ and arrive at an expression for the free energy per spin (see also \cite{binder-young})
\begin{equation}
f=-\frac\beta2(1-2q_1)+ \frac{\beta}4\sum_{i=1}^{\kappa+1} q_i^2(m_i-m_{i-1})-\tilde f
\end{equation}
with $\kappa$ the order of RSB and $q_i$ the values of the elements of the Parisi-blocks with size $m_i$ \footnote{We use the convention $q_i<q_{i-1}$ and $m_i<m_{i-1}$.}. The non-trivial part $\tilde f$ of the free energy is given by a $\kappa+1$-fold nested integral
\begin{equation}
\tilde f = \frac{T}{m_{\kappa}}\int_{\kappa+1}^G\log\left[\int^{GE}_\kappa \cdots \int_2^{GE}\int_1^G\left(2\cosh(\beta h_1)\right)^{m_1} \right]
\end{equation}
where two Gaussian integral operators are defined as
\begin{equation}
\int_i^G f(h_i)\equiv\int_{-\infty}^\infty\frac{dh_i}{\sqrt{2\pi \Delta q_i}} \exp\left(-\frac{(h_i-h_{i+1})^2}{2\Delta q_i}\right) f(h_i)
\end{equation}
with $\Delta q_i=q_{i}-q_{i+1},\sms \Delta q_{\kappa+1}=q_{\kappa+1}$ and
\begin{equation}
\int_i^{GE}f(h_i)\equiv\int_i^G f(h_i)^{r_{i-1}},\bs r_{i-1}=\frac{m_{i}}{m_{i-1}}.
\end{equation}
The last Gaussian integral in this chain is centered at the external field $h_{\kappa+2}=H$.

As explained above, it is convenient at low temperatures to transform to new variables $a_i=\beta m_i$. Since $r_i=\frac{m_{i+1}}{m_i}=\frac{a_{i+1}}{a_i}$ the definitions of the Gaussian integral operators $\int^G$ and $\int^{GE}$ remain invariant. The explicit temperature dependence is located at the initial condition, which has a well defined zero temperature limit in the $a$-formulation
\begin{equation}
\left(2\cosh(\beta h_1)\right)^{T a_1} \sms \overset{T\rightarrow 0}\longrightarrow \sms e^{a_1|h_1|}.
\end{equation}
The utility of our new formulation for low temperatures lies in the non-singular limit of this inner integrand. It has been shown before\cite{oppermann-prb2000} that the first integral can be performed analytically in the $T\rightarrow0$ limit. For finite temperatures, there is also an asymptotic regime $h_1\rightarrow\pm\infty$ of the integrand where
\begin{equation}
\left(2\cosh(\beta h_1)\right)^{T a_1} \sms \overset{h_1\rightarrow\pm\infty}\longrightarrow \sms e^{a_1|h_1|}.
\end{equation}
In the asymptotic regime, the Gaussian integrals can be performed analytically at each level of integration ($|h_{i+1}|\gg 1$) 
\begin{equation}
\int_i^{GE}e^{a_{i-1}|h_i|} \simeq \exp\left(a_i\left(|h_{i+1}|+\frac12a_i\Delta q_i\right) \right),
\end{equation}
while the deviation of the integrals from its asymptotic form at small $h$ is incorporated in form of auxiliary functions $\expc_i(h)$ defined by
\begin{multline}
\int_i^{GE}\cdots\int_1^G \left(2\cosh(\beta h_1)\right)^{T a_1} = \\ \exp\left(a_i\left(\frac12\sum_{j=1}^i a_i\Delta q_i+ |h_{i+1}|+\expc_i(h_{i+1}) \right) \right).
\end{multline}
The functions $\expc_i$ are well defined and free of singularities at each level of integration, for all temperatures including $T=0$ directly and at all $\kappa<\infty$. It is therefore the object of our choice for the finite RSB numerics. At zeroth level of integration the 'initial condition'
\begin{equation}
\expc_0(h)=T\log\left(1+\exp\left(-\frac{2|h|}{T}\right)\right)\label{initcond}
\end{equation}
has a smooth transition from finite temperatures to $T=0$, i.e. $\lim_{T\rightarrow0}\expc_0(h)=0$. Furthermore, the $\expc$ obeys a $h\rightarrow-h$ symmetry at each level of integration - irrespective of temperature $T$ or magnetic field $H$ and is continuous at $h=0$, though there is a kink at $h=0$ with $\left. \partial_h \expc_i(h)\right|_{h=0^+}=-1,\sms\forall i$. Therefore, it is sufficient to restrict ourselves to $h>0$ and introduce a boundary condition at $h=0$ in the continuous theory.

At finite $\kappa$, there are simple recursion relations between $\expc_i(h)$ and $\expc_{i-1}(h)$
\begin{widetext}
\begin{equation}
\expc_{i}(h)=\frac1{a_i}\log\left[\int^G_i \exp\left(a_i\left(|h'|-|h|+\expc_{i-1}(h') -\frac12 a_i\Delta q_i\right) \right) \right]\label{expcrecursion}
\end{equation}
\end{widetext}
and in the end, the non-trivial part of the free energy can be expressed in terms of an integral of $\expc_\kappa$
\begin{equation}
\tilde f=\int_{\kappa+1}^G(\expc_\kappa(h)+|h|) + \frac12 \sum_{k=1}^\kappa a_k\Delta q_k
\end{equation}
so the free energy per particle can be written as
\begin{multline}
f = -\frac T4\chi_{ne}^2 +\frac14 \sum_{i=1}^\kappa a_i\left[(q_i-1)^2-(q_{i+1}-1)^2\right] \\
- \int_{\kappa+1}^G(\expc_\kappa(h)+|h|) \label{freeenergy}
\end{multline}
with the non-equilibrium susceptibility $\chi_{ne} = \beta(1-q_1)$. For the continuous RSB limit, one can fix the domain boundaries of $q(a,T)$ by defining $a_{\kappa+1}\equiv 0$ and $a_0\equiv \beta$ and the first line of (\ref{freeenergy}) becomes proportional to an integral $\int_0^\beta da (1-q(a))^2$.

At finite $\kappa$, the free energy must be maximized with respect to $\{a_i,q_i\}$. This is done for $T=0$ (finite $T$) by a root search of the gradient of $f$ in the $2\kappa$ ($2\kappa+1$) dimensional parameter space \footnote{At zero temperature, $q_1=1$, because $q_1$ is the Edwards-Anderson order parameter.}.

\begin{figure}
\centering
\includegraphics[width=200pt]{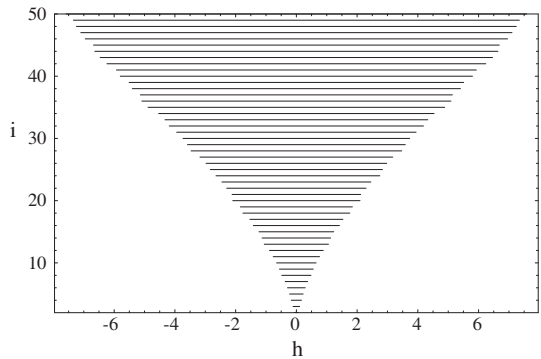}
\caption{Non-trivial domain of $\expc_i(h)$ of a typical 50-RSB calculation at $T=0$ and $H=0$ represented by the bars in $h$ direction. The non-trivial domain is defined as the interval where $\expc_i(h)$ is numerically greater than zero.}
\label{nontrivdomain}
\end{figure}

\begin{figure}
\centering
\includegraphics[width=250pt]{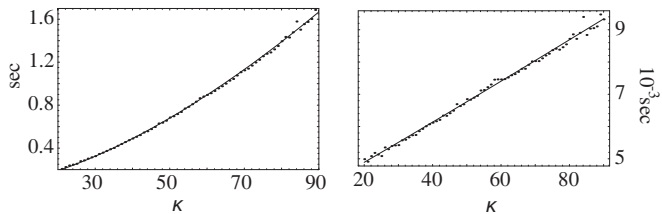}
\caption{Calculation times at $T=0$ in dependence of $\kappa$. The calculation time of the free energy with all derivatives (left part) goes with $\kappa^{1.6}$ and the calculation time of a single derivative (right part) goes with $\kappa^{1.1}$}
\label{calctimes}
\end{figure}

There are two main advantages of performing numerics in terms of $\expc$. The first is that the non-trivial domain of this function (i.e. where it is non-zero, see Fig. \ref{nontrivdomain}) is strongly restricted to the region of small $h$ and so the numerical effort reduces considerably. This is illustrated by Fig. \ref{calctimes} where the $\kappa$-dependence of typical calculation times is shown. A single calculation implies the calculation of the free energy and all its derivatives. Since the number of derivatives grows as $2\kappa+1$, the full calculation time grows stronger than linear. The calculation time normalized to the number of derivatives to calculate, however grows nearly linearly with $\kappa$ and therefore high orders of RSB may be obtained. The second advantage is, that $\expc(h)\sim \mathcal O(1)$ and no loss of numerical precision due to large numbers happens.

\begin{figure}
\centering
\includegraphics[width=210pt]{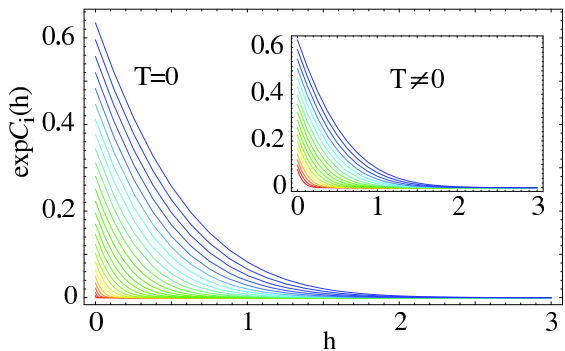}
\caption{(color-online): $\expc_i(h)$ at even levels of integrations at $T=0$ and $\kappa=50$. The lower lines (red) correspond to small $i$ and the upper lines (blue) to large $i$. The inset shows a $T=0.1$ calculation with $\kappa=30$.}
\label{expcFigure}
\end{figure}

Figure \ref{expcFigure} shows the functions $\expc_i$ at all integration levels in case of zero temperature and for finite temperatures. The initial condition is represented by the lowest curve, which is exactly zero for $T=0$ and remains finite for finite temperature corresponding to the non-zero initial condition given in equation (\ref{initcond}).

\section{Results}
In this section, we discuss some results of our extensive numerical computations according to the scheme described in section III. We have been able to perform calculations at extremely high orders of RSB (up to 200 at $T=0$ and up to 53 at finite temperatures) with unprecedented accuracy which allow a deep insight into the subtleties that arise as $\kappa$ approaches its physical limit $\kappa=\infty$.

In a previous publication we used 42 RSB data to extrapolate the zero temperature free energy to $\kappa=\infty$ by fitting $f_\kappa$ to the function $f_\kappa = f_{\kappa=\infty} + \frac{const.}{(\kappa+\kappa_0)^\alpha}$\footnote{$\kappa_0\simeq1.27$ is an offset which is used to improve the fit quality.}. The extrapolation is now confirmed according to 200 RSB high precision numerics with the new $\expc$-formalism to $f_{\kappa=\infty}= -0.763\;166\;726\;566\;547$, $\alpha=4$. It is difficult to estimate the error of the extrapolated free energy. The fluctuations of the free energy of the finite $\kappa$ calculations are on an arbitrary small scale, due to the utilization of arbitrary precision arithmetics, so that the accuracy of $f_{\infty}$ depends only on the fit procedure. To give the reader an idea of the convergence level at 200 RSB, let us state that $f_{200} - f_{\infty} \simeq 10^{-11}$. With standard fit methods, an accuracy of $10^{-13}$ is obtained, which represents the minimum accuracy of $f_{\infty}$. With some technical tricks which are beyond the scope of this paper, however, it seems that the accuracy might be up to $10^{-15}$. In any case, this is by far the most precise ground state energy ever obtained for the SK-model and it provides a good test for all coming formalisms which work directly at $\kappa=\infty$ and $T=0$. In comparison with the literature, we find that the value is consistent with the estimate of Parisi \cite{Parisi-T0}, but is slightly outside of the confidence interval given in \cite{crisanti-rizzo}.

At zero temperature the free energy $f$ equals the internal energy $u$. At finite temperatures, they are still closely related. We have checked numerically, that the free energy can be expanded in a Taylor series with regular exponents at $T=0$
\begin{equation}
f(T) = f_0- s_0T+f_2 T^2+f_3 T^3 + \mathcal O(T^4)
\end{equation}
where $s_0 = -\chi_{ne}^2/4$ is the zero temperature entropy, which can also be expressed in terms of the non-equilibrium, or single-valley susceptibility $\chi_{ne}$. The internal energy can be expressed by the same coefficients $u(T)=f - T\frac{d f}{dT}  = f_0- f_2 T^2- 2f_3 T^3 +\mathcal O(T^4)$, and also the temperature dependence of the entropy can be written as $s(T)=s_0 - 2f_2 T -3 f_3 T^2 +\mathcal O(T^4)$. From our results we find $f_2\rightarrow 0$ and $f_3\rightarrow -0.24$, in agreement with the literature \cite{Pankov,crisanti-rizzo}. 

The zero temperature entropy is directly related to the non-equilibrium susceptibility which can be written at finite RSB as \footnote{Alternatively, it is also obtained by varying the free energy with respect to $q_1$. This has been checked and yields the same result.}
\begin{equation}
\chi_{ne} = -\frac12 \sum_{i=1}^\kappa a_i(q_i^2-q_{i+1}^2)-f.
\end{equation}
One finds, that $\chi_{ne}\propto(\kappa+\kappa_0)^{-\gamma}$ with $\gamma=1.666664\pm 5 \cdot 10^{-6}$. For an exponent, this numerical accuracy is sufficient to claim $\gamma=5/3$. As a result, both, the non-equilibrium susceptibility and the entropy vanish with irregular exponents $5/3$ and $10/3$ for $\kappa\rightarrow\infty$ at zero temperature. The implications of those irregular exponents will be discussed elsewhere \cite{oppermann-scaling}.

\begin{figure}
\centering
\includegraphics[width=200pt]{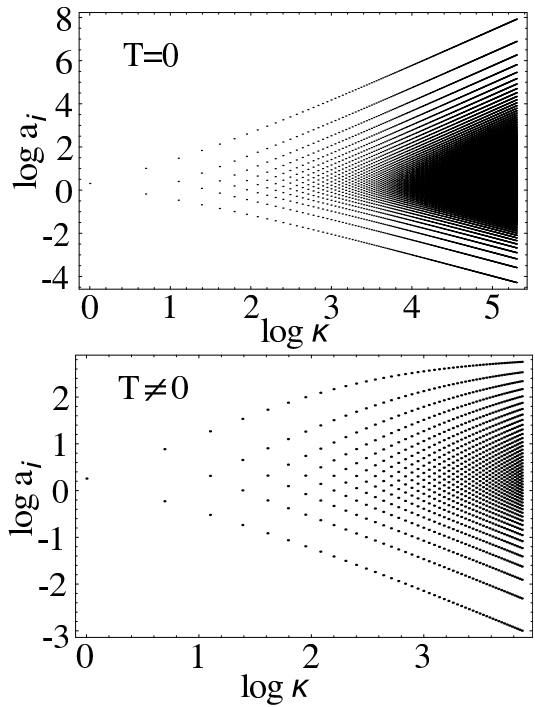}
\caption{$\log a_i$ as a function of $\log \kappa$ for $\kappa=1,..,200$ at $T=0$ and for $\kappa=1,..,50$ for $T=0.03$.}
\label{alevels}
\end{figure}

As we have explained above, the natural formulation of the order function at low temperatures is in terms of a function $q(a,T)$ since it resolves the structure at the singular point $x=0$ of the original Parisi order function $\tilde q(x,T)$ at $T=0$. In order to better understand the critical properties of the zero temperature order function, we have to discuss the finite RSB step approximation to $q(a,T)$. In Fig. \ref{alevels} the logarithm of the Parameters $a_i$ are plotted as a function of $\log\kappa$. One can clearly see that for large $a_i$ the spacing between successive $a$-points does not vanish, while for moderate $a_i$, a continuum emerges. Also at small $a_i$, a spacing appears on a $\log a$-scale. This small-$\log a$ spacing, however, does not imply a spacing on an $a$ scale, where the differential equations of a continuous theory appear. At finite temperatures, the discreteness at large $a$ disappears due to the restrictedness of $a_i$ to the interval $[0,\beta]$, while the small $a$ discreteness remains. A finite external field would destroy also the small-$a$ discreteness. The notion of non-zero plateau-widths on a log scale is directly related to the discrete spectra of the $r_i$-levels, which will be discussed more thoroughly elsewhere \cite{oppermann-scaling}.

At $T>0$, the identification of a break point ${\bar x}$ is important. One finds, that 
\begin{equation}
{\bar x}=\lim_{\kappa\rightarrow\infty} T a_1. 
\end{equation}
From our computations at finite temperatures, we can extract a confident value for ${\bar x}$ at temperatures down to $T=0.015$. Below this temperature, calculations at $\kappa>50$ are needed. There is, however, no reason to expect a large deviation of ${\bar x}(T=0.015)$ from its true extrapolation to $T=0$. We find ${\bar x}(T=0.015) = 0.54684 \pm 2\cdot 10^{-5}$, in consistence with the literature \cite{Pankov,crisanti-rizzo}. At zero temperature, ${\bar x}$ is ill defined because $\tilde q(x,0) = 1,\;\forall x>0$. As a consequence, one does not find ${\bar x}$ directly in the $T=0$-theory. In fact, there is a subtle non-commutativity of the $T\rightarrow0$ and $\kappa\rightarrow\infty$ limits for quantities like the break-point \cite{oppermann-scaling}.

At finite $\kappa$ and $T=0$, no $m_i$ remains finite, all go to zero. From scaling arguments, however, one can see that there are finite $m_i$ even at $T=0$ in the $\kappa=\infty$ limit. These are exactly the $m_i=T a_i$ corresponding to the discrete $a$-points (see Fig. \ref{alevels}) and so the finite $m_i$ are also discrete \footnote{This can also be seen from the discrete spectra, because the definition of $r_i=a_i/a_{i-1} = m_i/m_{i-1}$}. In the continuous formulation of the infinite RSB limit, an initial condition of a partial differential equation is given at $m_1$ \footnote{Sometimes in the literature one finds the initial condition at $x=1$, which is equivalent due to the triviality of the differential equations beyond the break point.}. There is, however, the region of discrete $m_i$ where a continuous theory is invalid, so at $T=0$ the initial condition of the continuous theory is disconnected from the validity domain of the differential equation. This is why the traditional theory fails for $T=0$.

Two other important and closely related parameters are the $T^2$-coefficient of the $T$-expansion of the Edwards-Anderson order parameter $q_{EA}(T)=1-\alpha T^2+\mathcal O(T^3)$ and the $a^{-2}$-coefficient of the $1/a$-expansion of $q(a,0) = 1-\gamma a^{-2}+\mathcal O(a^{-3})$. At finite orders of RSB, there is also a linear term in the $T$-expansion of $q_{EA}$ with the coefficient equal to $\chi_{ne}$, but this coefficient vanishes for $\kappa\rightarrow\infty$. From our numerical data, we can extract $\alpha=1.59411\pm0.00002$ and $\gamma=0.4108\pm0.0001$.

\begin{figure}[here]
\centering
\includegraphics[width=220pt]{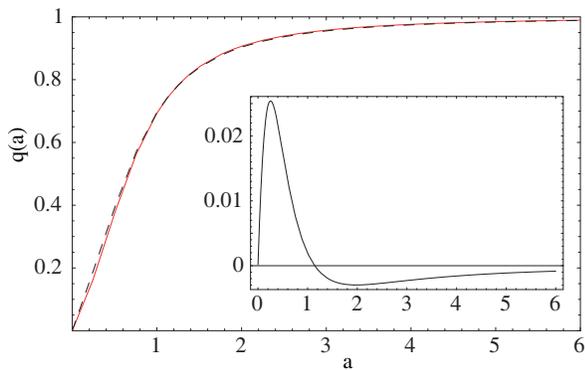}
\caption{(color-online): Comparison of the universal scaling function obtained by the PaT-hypothesis (black, dashed) and the zero temperature 200 RSB order function $q(a,0)$ (red, solid). The inset shows the difference of the two curves.}
\label{patcomp}
\end{figure}

In the following paragraph, we analyze our results from the viewpoint of PaT scaling \cite{PaT-scaling}. In the original formulation, PaT scaling has been used to evaluate a (approximately) universal scaling function $\tilde q(x,T)\simeq f(x/T)$ for $x<{\bar x}$. This scaling is known to not hold exactly. If it held exactly, then $f(x/T)=q(x/T,0)$ at all temperatures. In Fig. \ref{patcomp} we compare $q(a,0)$ to the result of the PaT hypothesis, which was obtained according to the description in \cite{PaT-scaling}.

In order to discuss the correction to PaT scaling near $T=0$, we write
\begin{equation}
q(a,T) = \left\{\begin{matrix} 
q(a,0) + \tilde q(a,T) & \text{ for } & a<\bar a \\
q(\bar a,0) + \tilde q(\bar a,T) & \text{ for } & a\geq \bar a
                \end{matrix}\right.
\end{equation}
with $\bar a =\beta \bar x$ and $\tilde q(a,T)$ is the correction to PaT scaling near $T=0$. From the identity $1=\int_0^{\beta} da (1-q(a,T))$ one can derive the correction to the PaT-scaling break point at zero temperature
\begin{equation}
{\bar x}(T=0) = \frac12 - \frac 1{2\alpha} \lim_{T\rightarrow0}  \int_0^{\bar a} da \; \frac d{dT} \tilde q(a,T) \label{bpcorrection}
\end{equation}
with $\alpha$ the quadratic temperature coefficient of the Edwards-Anderson (EA) order parameter $q_{EA}(T) \simeq 1-\alpha T^2$. If we assume that $\tilde q(a,T)$ can be expanded in a Taylor-series near $(a,T)=(\infty,0)$, i.e.
\begin{equation}
\tilde q(a,T) = \sum_{i} T^i \; \tilde q_i(a),\bs \text{for } T\ll 1
\end{equation}
and
\begin{equation}
\tilde q_i(a) = \sum_j b_i^j a^{-j},\bs\text{for } a\gg 1
\end{equation}
one can derive relations between the lowest coefficients of the $\tilde q(a,T)$ expansion, the quadratic temperature coefficient $\alpha$ of the EA order parameter, the break point and the $a^{-2}$ coefficient $\gamma$ of the expansion of $q(a,0)$ at $a=\infty$.
\begin{eqnarray}
b_1^0 = b_1^1 &=&0 \\
\alpha &=& \frac{\gamma}{{\bar x}^2} - b_2^0\\
\int_0^\infty da\; \tilde q_1(a)  &=& \frac\gamma{{\bar x}^2}(1-2 {\bar x}) - b_2^0.\label{intrel}
\end{eqnarray}
The first relation states that $\tilde q_1(a)$ must go to zero faster than $a^{-1}$ as $a\rightarrow\infty$. The parameters $\gamma,\alpha,\bar x$ are very well known from the literature and have been obtained from our numerical data, too. From the above relation, one can thus extract $b_2^0=-0.22035\pm0.00012$ and write the correction to PaT-scaling near $(a,T)=(\infty,0)$ as
\begin{equation}
q(a,T) = q(a,0) -0.22 \; T^2 + \mathcal O(T,a^{-1})^3.
\end{equation}
Relation (\ref{intrel}) can be used as a check for $\tilde q_1(a)$ extracted from numerics.

\begin{figure*}
\centering
\includegraphics[width=350pt]{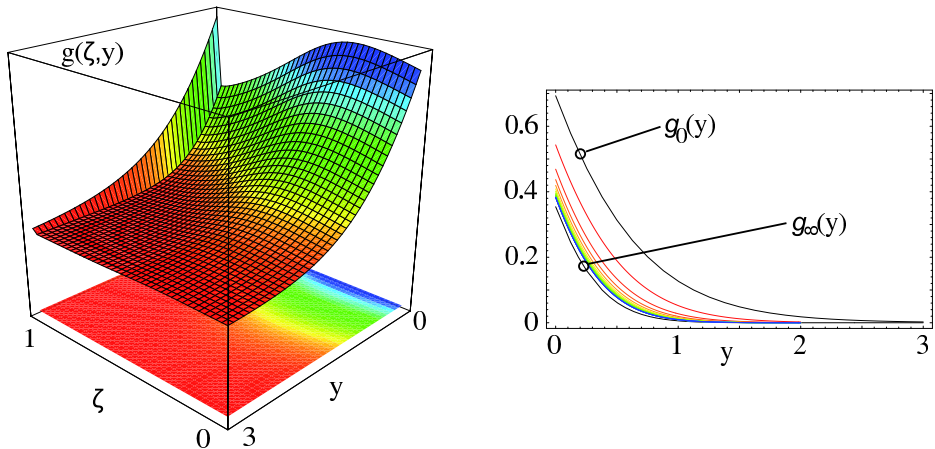}
\caption{(color-online): Left: $g(\zeta,y)$ with $\zeta=a/(1+a)$. Right: The initial condition $g_0(y)$ (top black line) at $x=1,a=\infty$ and the first 20 integrations (red to blue). The bottom black line shows the initial condition $g_{\infty}(y)$ at $x=0,a=\infty$. }
\label{erfuncSing}
\end{figure*}

\section{$\infty$-RSB limit}
As $\kappa\rightarrow\infty$, the number of parameters $(q_i,a_i)$ goes to infinity and a smooth order function $q(a,T)$, defined on $[0,\beta]$ rises for $T>0$. In this limit, the set of functions $\expc_i$ become one continuous function of the variables $a$ and $h$ defined as
\begin{equation}
\expc(a_i,h)=\lim_{\kappa\rightarrow\infty} \expc_i(h)
\end{equation}
and the free energy for $H=0$ can be written as
\begin{equation}
f=-\frac14\int_{0}^\beta da (q(a)-1)^2-\expc(a=0,h=0)
\end{equation}
where $\expc(a,h)$ is obtained by solving the partial differential equation
\begin{equation}
\partial_a \expc=-\frac{\dot q(a)}2\left[\partial_{h}^2\expc+2a \partial_h\expc+a\left(\partial_h\expc\right)^2\right] \label{expcdgl}
\end{equation}
with boundary and initial conditions
\begin{eqnarray}
\left. \partial_h\expc(a,h)\right|_{h=0^+}=-1;\sms\expc(a,\infty)=0 \nonumber \\
 \expc(\beta ,h)= T\log\left(1+\exp\left(-2\beta h\right)\right) \label{expcinit}
\end{eqnarray}
for $h\geq 0$.

\begin{figure}[here]
\centering
\includegraphics[width=240pt]{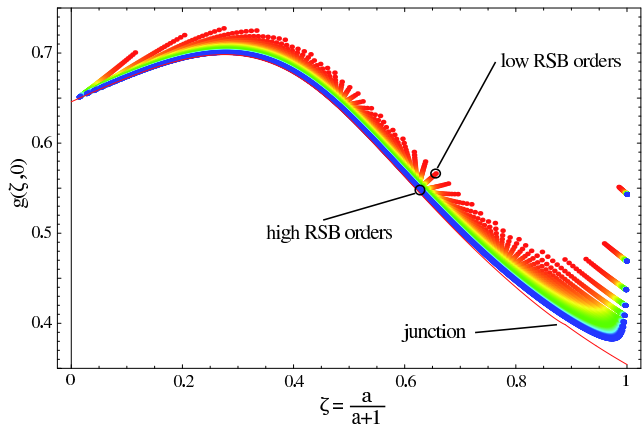}
\caption{(color-online): $g(\zeta,0)$ with $\zeta=a/(1+a)$ for up to 200 orders of RSB. The dots with the different colors refer to low (red) and high (blue) RSB orders. The solid line (red) is obtained by the $\infty$ RSB formalism proposed in the text.}
\label{erfuncInfty}
\end{figure}

At finite temperatures, where the spacing between successive $a_i$ and $q_i$ go to zero as $\kappa\rightarrow\infty$, the $\expc$ is well-behaved and the differential equation can be solved numerically. In this case equations (\ref{expcdgl},\ref{expcinit}) are merely a reformulation of the Parisi theory \cite{ParisiRSB,binder-young}, convenient for investigations at low temperatures. At exactly zero temperature, however, two problems arise which force us again to reformulate. Firstly the discreteness in the large $a$ regime of $q(a)$ formally invalidate the differential equation at $a=\infty$, where the initial condition is given. Secondly the second derivative of the initial condition has a divergence at $h=0$ as $T\rightarrow0$.

The second issue can be resolved by a further rescaling $y = (a+1) h$ and introduction of the function $g(a,y) = (a+1)\expc(a,y/(a+1))$. The initial condition at $\beta$ for $\expc$ translates to $g(\beta,y) = \log(1+e^{-2y}) = g_0(y)$. The left part of Fig. \ref{erfuncSing} shows the $a$ and $y$-dependence of $g(a,y)$ (for visualization purposes, the $a$-axis has been rescaled to the interval [0,1] by the introduction of the variable $\zeta=a/(1+a)$). It shows that for finite $a$ (i.e. $\zeta<1$), $g(a,y)$ varies smoothly as a function of $a$ and the description in terms of a partial differential equation is valid. In the large $a$ limit, however, a singularity appears. To understand this singularity and the solution of this issue, we go back again to the finite RSB formulation. The investigation of the 200 RSB calculations shows that at large $a$, the difference between $g(a_i,y)$ and $g(a_{i+1},y)$ does not vanish with $\kappa\rightarrow\infty$ as it does for moderate $a\ll a_1$ \footnote{$a_1$ is the largest $a$-parameter in the finite RSB-formulation} (see right side of Fig. \ref{erfuncSing}). This means, that the recursion relation (\ref{expcrecursion}) does not approach a differential equation at $a=\infty$. Instead, one can see that the recursion drives the function to a limiting function $g_\infty(y)$ from which the continuous part of $g(a,y)$ starts. In the right part of figure \ref{erfuncSing} the first 10 results of the recursion relation for a $\kappa=200$ calculation are shown together with $g_0(y)$ and $g_\infty(y)$, which is obtained by numerically solving an ordinary differential equation, as explained below.

To complete our discussion, we now approach the point $a=\infty$ from below. In the finite $a$ regime, the behavior of $g(a,y)$ is governed by the partial differential equation (the dot refers to a derivative with respect to $a$, while the prime means a $y$-differentiation)
\begin{equation}
\dot g = -\frac{\dot q}2(a+1)\left[(a+1) g'' + 2 a g' + a (g')^2\right] + \frac{g-y\; g'}{a+1}.\label{gdgl}
\end{equation}
In order to investigate this equation at zero temperature in the limit $a\rightarrow\infty$, where the initial condition is given, we expand the order function $q(a,0) = 1 - \gamma a^{-2}$ and equation (\ref{gdgl}) itself near $a=\infty$. To first order in $a^{-1}$ we find $\dot g = \frac1a F[g]$ with
\begin{equation}
 F[g]=g-y\; g'-\gamma (g''+2g'+(g')^2).
\end{equation}
Obviously, an initial condition $g(\infty,y)$ of (\ref{gdgl}) with $F[g(\infty,y)]\neq 0$ would lead to a logarithmic singularity of $g(a,y)$ at $a=\infty$. The only non-singular initial condition is therefore the solution to the ordinary differential equation $F[\tilde g_\infty(y)]=0$. Indeed, the solution $\tilde g(y)$ of this differential equation seems to be the limiting function $g_\infty(y)$ of the recursion starting from $g_0(y)$ discussed above. In Fig. \ref{erfuncSing}, the function $g_\infty(y)$ is the numerical solution of $F[g_\infty]=0$. In some sense, the partial differential equation governing the function $g(a,y)$ yields its own initial condition - it is the only initial condition which makes sense.

To further illustrate and confirm this line of reasoning and to better understand the transition $\kappa\rightarrow\infty$ we shall restrict our discussion of $g(a,y)$ to $y=0$ as representative for the $a$-dependence of $g$. In Fig. \ref{erfuncInfty} we plot $g(a_i,0)$ at $T=0$ for $\kappa<\infty$ varying from 10 to 200. Again, one can see the discreteness at $a=\infty$. For demonstration purposes, a numerical solution of (\ref{gdgl}) is plotted as the drawn through line. To obtain this solution at large $a$, equation (\ref{gdgl}) has been expanded up to order $a^{-2}$ at $a=\infty$. The solution of this expansion taken at $a=8$ ($\zeta\simeq0.89$) is then used as an initial condition for the full partial differential equation (\ref{gdgl}) at $a=8$. By thoroughly looking at the line, one can see a small error near this junction point. With more effort like higher order expansions at $a=\infty$ or advanced numerical methods for partial differential equation (e.g. pseudo-spectral methods), the quality of the full continuous RSB solution at zero temperature can be strongly improved, but this is beyond the scope of this work.

\section{Conclusion}
We have developed an RSB technique which allows calculations at extremely high orders of RSB near and at $T=0$. We have indeed performed calculations at $T=0$ for up to 200 RSB and for finite temperatures up to 53 RSB. We extracted the dependence of various observables on the order of RSB and on temperature and obtained, to our best knowledge, the by far most precise numerical value for the ground state energy of the SK model. The connection to PaT-scaling has been discussed and a correction to the PaT-scaling assumption $q(x,T)=f(x/T)$ near zero temperature and $a=x/T=\infty$ has also been discussed. Furthermore, we have proposed an Ansatz for a full treatment of the zero temperature limit of the SK-model directly at infinite RSB - in analogy to the continuous RSB formalism at finite temperatures. It would be usefull to derive a closed set of self-consistency equations for $q(a)$ in the sense of Sommers and Dupont \cite{Sommers-Dupont} and solve them numerically in order to obtain the zero temperature order function directly in its physical limit.

We gratefully acknowledge useful discussions with David Sherrington, Kay Wiese and Markus Müller.

\end{document}